# Suppression of star formation in dwarf galaxies by grain photoelectric feedback

John C. Forbes[1], Mark R. Krumholz[1,2], Nathan J. Goldbaum[3] & Avishai Dekel[4]

[1] Department of Astronomy & Astrophysics, University of California, Santa Cruz, CA 95064

[2] Research School of Astronomy & Astrophysics, Australian National University, Canberra, ACT 2601, Australia 2611

[3] National Center for Supercomputing Applications, University of Illinois at Urbana-Champaign, 1205 W. Clark Street, Urbana, IL 61801

[4] Center for Astrophysics and Planetary Sciences, Racah Institute of Physics, The Hebrew University, Jerusalem 91904, Israel

**Photoelectric heating has long been recognized as the primary source of heating for the neutral interstellar medium[1]. Simulations of spiral galaxies[2] found some indication that photoelectric heating could suppress star formation. However, simulations that include photoelectric heating have typically found that it has little effect on the rate of star formation in either spiral galaxies[3,4] or dwarfs[5] suggesting that supernovae and not photoelectric heating are responsible for setting the star formation law in galaxies[6-8]. This result is in tension with recent work[9-13] indicating that a star formation law that depends on galaxy metallicity, as expected for photoelectric heating but not for supernovae, reproduces the present-day galaxy population better than a metallicity-independent one. Here we report a series of simulations of dwarf galaxies, where the effects of both photoelectric heating and supernovae are expected to be strongest. We simultaneously include space- and time-dependent photoelectric heating, and we resolve the Sedov phase of every supernova blast wave, allowing us to make a direct measurement of the relative importance of momentum injection by supernovae and dust heating by far ultraviolet (FUV) photons in suppressing star formation. We find that supernovae are unable to account for the long observed[14] gas depletion times in dwarf galaxies. Instead, ordinary photoelectric heating is the dominant means by which dwarf galaxies regulate their star formation rate at any given time, suppressing the star formation rate by more than an order of magnitude relative to simulations with only supernovae.**

To investigate whether the depletion times in dwarf galaxies, which are longer than for Milky Way-like galaxies by more than an order of magnitude[15,14], are set by the momentum injection from supernovae or by photoelectric heating, we perform a series of high-resolution hydrodynamic simulations using the Enzo adaptive mesh refinement code[16]. We include a new prescription for supernova and pre-supernova stellar feedback, and a new method for self-consistent spatially-dependent photoelectric heating (see Methods). We use two sets of initial conditions. Both correspond to isolated dwarf galaxies with an initially laminar exponential disk, a stationary hot halo, and collisionless particles representing stars and dark matter. The galaxies have a dark matter halo mass of $10^{10}$ $M_\odot$, a stellar mass of $10^7$ $M_\odot$, and an observationally-motivated[17] cold gas mass of $10^8$ $M_\odot$ Galaxies in this mass range are comfortably above the limit where star formation can be quenched by the cosmological UV background[18], but small enough that the effects of both heating by FUV photons and supernova feedback[19] are plausibly extreme. The initial conditions differ in the exponential scale length chosen for the gas; one set uses 5 kpc, designed to mimic recently-discovered nearly-starless galaxies[20], and toward the high end of the range observed for field dwarf galaxies[14]. The other set uses 1 kpc, toward the low end of the observed field dwarf range. For a galaxy with an HI mass of $10^8$ $M_\odot$, assuming an exponential HI profile, the observed relation[21] between HI mass and HI size suggests an HI scale length of about 1.9 kpc.

To understand how supernovae and photoelectric heating each contribute to the evolution of these galaxies, we perform a straightforward numerical experiment. We run a fiducial simulation of the 5 kpc scale length initial conditions including both supernovae and photoelectric heating, and simulations where each of these effects is turned off in turn. We refer to these as the "SN+PE", "PE Only," "SN only," and the "No feedback" simulations. We also run the "SN+PE", "PE Only", and "No feedback" cases for the 1 kpc initial conditions. The simulations with supernovae also include pre-supernova stellar feedback from winds and HII regions. For the 5 kpc case, we re-run with the four different feedback models at three different maximum spatial resolutions -- 10 pc, 5 pc, and 2.5 pc. These resolutions are high enough, and the typical densities in which supernovae explode in these simulations are low enough, that our simulations do not suffer from the overcooling problem[22], whereby poorly-resolved simulations overestimate the rate at which SN-heated gas cools (see Extended Data Fig. 1). We focus first on the 10 pc resolution simulations for the 5 kpc scale length initial conditions, since we have run these for the longest time. In the Methods section we compare to the higher resolution runs to evaluate the level of convergence of our results. We compare the 5 kpc and 1 kpc initial conditions below.

We find that all of the simulations follow a similar initial transient behavior. The gas disk cools from its center outwards, causing the disk to collapse vertically. Stars form first in the center, then further and further out. The central region of the galaxy after 90 Myr is shown, with the aid of the yt[23] package, in Figure 1 for each of the four feedback models. In terms of large-scale morphology, the supernovae have the most dramatic effect, driving large outflows with mass loading factors of order 100. Photoelectric heating slightly alters the global structure of the gas, but the two simulations without supernovae look quite similar.

Figure 2 shows the star formation rates and depletion times as a function of time for each of the four feedback models. After each simulated galaxy experiences its initial transient as the gas collapses from its initial state, their instantaneous star formation rates and depletion times are strikingly different depending on whether photoelectric heating was included. However, if two simulations are different only in their inclusion of supernova feedback, they end up with similar depletion times. This immediately shows that photoelectric heating, not feedback from supernovae, is primarily responsible for the long depletion times observed in dwarf galaxies. Supernovae, or even a lack of feedback, can result in low star formation rates in the long run by rapidly ejecting gas or locking gas into stellar remnants. This can be seen in the central kpc of the simulations without photoelectric heating - the star formation rate falls in this region, but the depletion time is relatively unaffected. Only the simulations including photoelectric heating produce depletion times in reasonable agreement with the large values frequently observed in dwarfs, as illustrated in Extended Data Fig. 2.

Do these results depend on the gas scale length? In Figure 2, the thin lines show the results for our simulations with a 1 kpc gas scale length. We find that the more compact galaxies have higher star formation rates and shorter depletion times, explaining the wide range of depletion times visible in the observational data. However, even for the 1 kpc simulations we find that, when supernovae are disabled but photoelectric heating is left active, the star formation rate is virtually unaffected. In contrast, disabling photoelectric feedback again causes the depletion time to drop by an order of magnitude, to values inconsistent with the observed sample. This indicates that photoelectric heating and not supernovae regulate star formation over a wide range of gas surface density in dwarf galaxies, and not just in the potentially-extreme[20,24] nearly starless dwarfs.

The means by which photoelectric heating suppresses star formation in our simulations is simple and intuitive. When a new star cluster is formed in the simulation, its most massive stars will emit photons with energies between 8 and 13.6 eV. These photons dominate the heating rate owing to the grain photoelectric effect, since they have energies high enough to liberate electrons from dust grains, but low enough not to be absorbed by intervening neutral hydrogen (Figure 3). The heating rate in the vicinity of a newly formed star increases the equilibrium temperature of the gas at fixed pressure or density. This in turn increases the Jeans mass of the gas and makes star formation more difficult (Figure 4).

We conclude that the physics responsible for setting the instantaneous star formation law in dwarf galaxies, i.e. the depletion time, is nearly independent of the physics determining the properties of the outflows. Ultimately young stars are responsible for both photoelectric heating and supernovae, but it is the moderate, local, instantaneous, volumetric heating of the former which controls the depletion time. Supernovae, despite their ability to eject mass from the galaxy at a rate larger than the star formation rate, are unable to shut down star formation locally and instantaneously. In the long run, even though dwarf galaxies have 10-100 Gyr depletion times, large mass loading factors mean that the timescale on which gas is lost from the ISM is shorter -- between 1 Gyr and 10 Gyr, implying that these galaxies may be in equilibrium between gas inflow and star formation plus outflows[25-27]. In the long run, therefore, the star formation rate in these galaxies would be set by the value of the mass loading factor, whereas the mass of gas in the ISM would be set by the depletion time. The former, in turn, is set by supernovae, and the latter by photoelectric heating.

**Acknowledgements** JCF and MRK acknowledge support from Hubble Archival Research grant HST-AR-13909. This work was also supported by NSF grants AST-09553300 and AST-1405962, NASA ATP grant NNX13AB84G, and NASA TCAN grant NNX14AB52G (JCF, MRK, and NJG), and by Australian Research Council grant DP160100695. NJG acknowledges additional support from ACI-1535651 and the Gordon and Betty Moore Foundation's Data-Driven Discovery Initiative through Grant GBMF 4651 to Matthew Turk. AD acknowledges support from the grants ISF 124/12, I-CORE Program of the PBC/ISF 1829/12, BSF 2014-273, and NSF AST-1405962. Simulations were carried out on NASA Pleiades and the UCSC supercomputer Hyades, supported by NSF grant AST-1229745.

**Author Contributions** JCF and NJG developed modifications to the publicly available Enzo code used in this work. The code was run and the results were analyzed by JCF. The manuscript was written by JCF and edited by all authors. The work was supervised and routinely advised by MRK and AD.

**Competing Interests** The authors declare that they have no competing financial interests.

**Correspondence** Correspondence and requests for materials should be addressed to JCF (email: jcforbes@ucsc.edu).

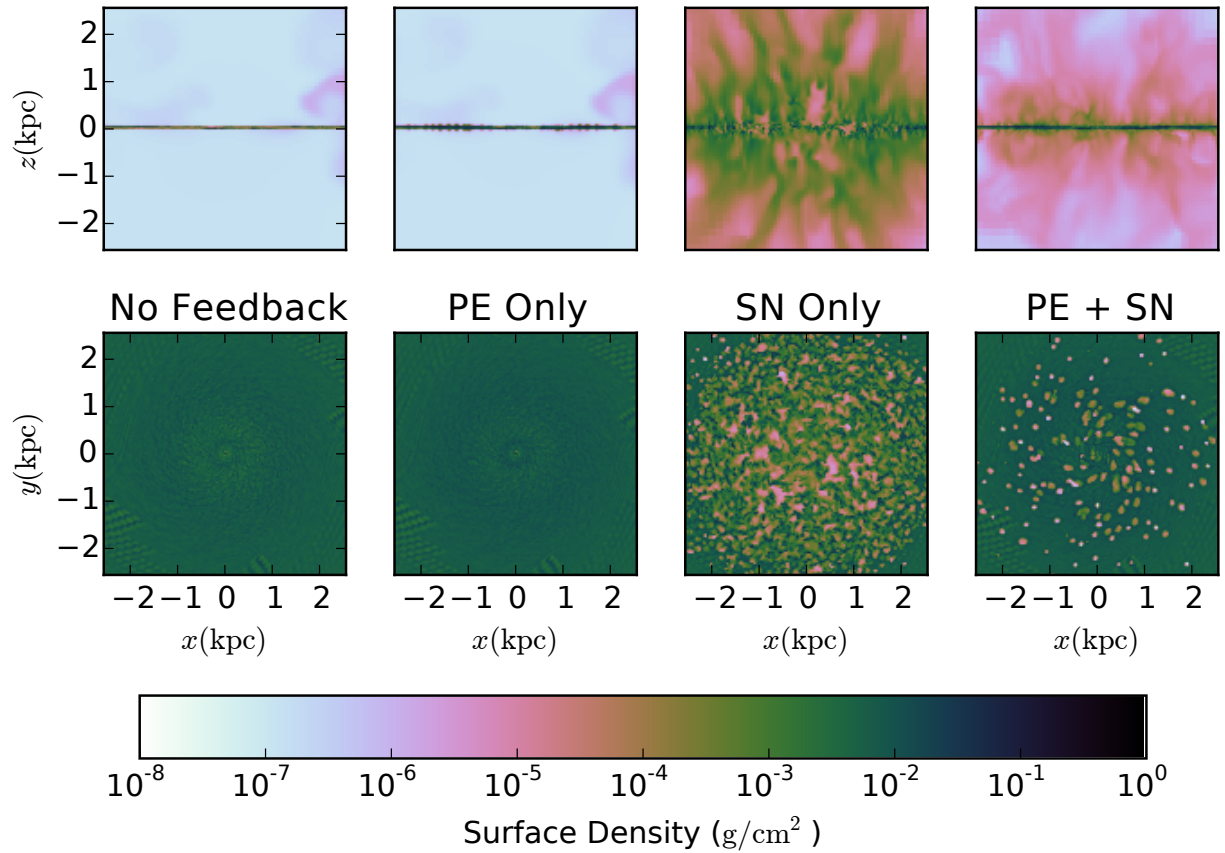

Figure 1. The morphology of the gas. For each of the 5 kpc scale length simulations after 90 Myr of evolution, the density is integrated ± 200 pc in the *y* dimension (top panels) and in the *z* dimension (bottom panels). The morphology of the disk is essentially determined by the presence of supernovae, despite the fact that the PE Only and PE+SN runs have nearly identical star formation rates (Figure 2). The SN Only simulation has an order of magnitude higher SFR than the PE+SN simulation, which is why the outflow and the disruption of the cold disk is more dramatic.

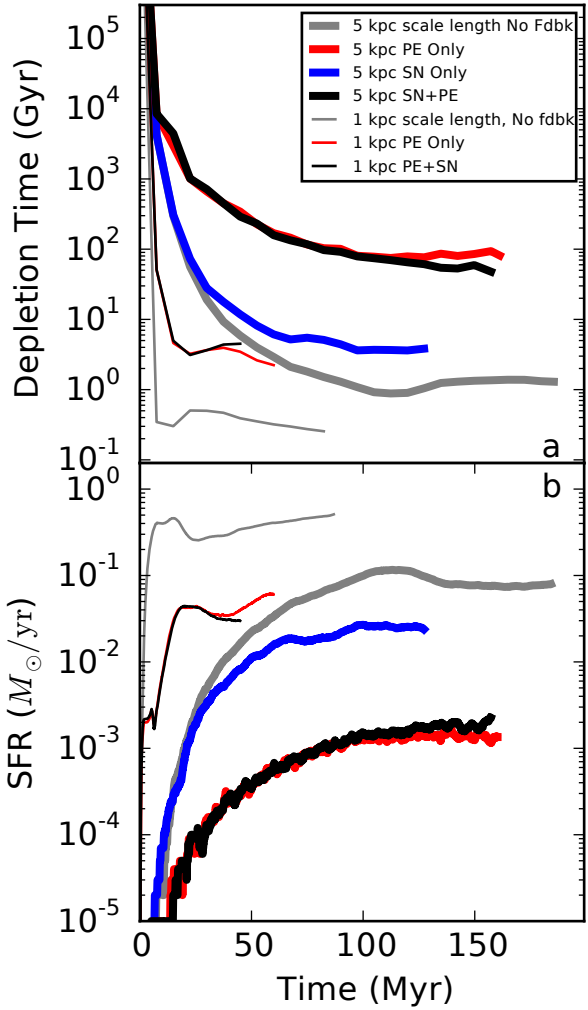

Figure 2. The star formation rates of the simulations. The long depletion times frequently observed in dwarfs (see Extended Data Fig. 2) are reproduced when both photoelectric heating and supernova feedback are included. Remarkably, turning off supernovae has almost no effect, meaning that photoelectric heating alone is responsible for the long depletion times in the simulations. More compact galaxies have higher star formation rates and shorter depletion times, but again disabling supernovae has a much smaller effect than disabling photoelectric heating.

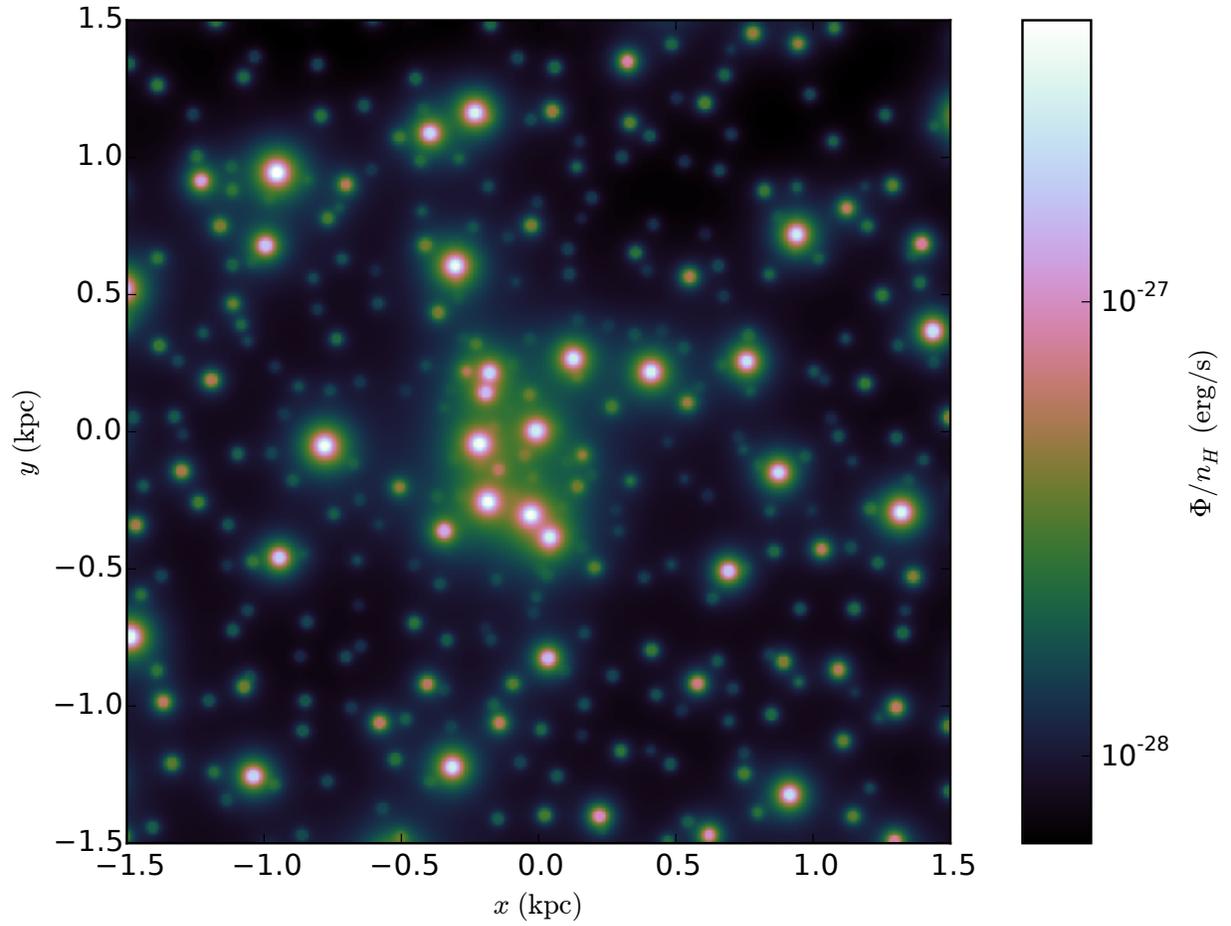

Figure 3. The photoelectric heating rate. Every young star particle in the simulation produces FUV radiation. The flux from all of these stars is summed in each cell (see Methods), yielding the volumetric heating rate shown here. Given the low surface density of young stars, this distribution is highly inhomogeneous. The low metallicity and small size of the galaxy means that the mean free path of the FUV photons is large, so the flux from each star and hence the heating rate falls off as $r^{-2}$ from each source.

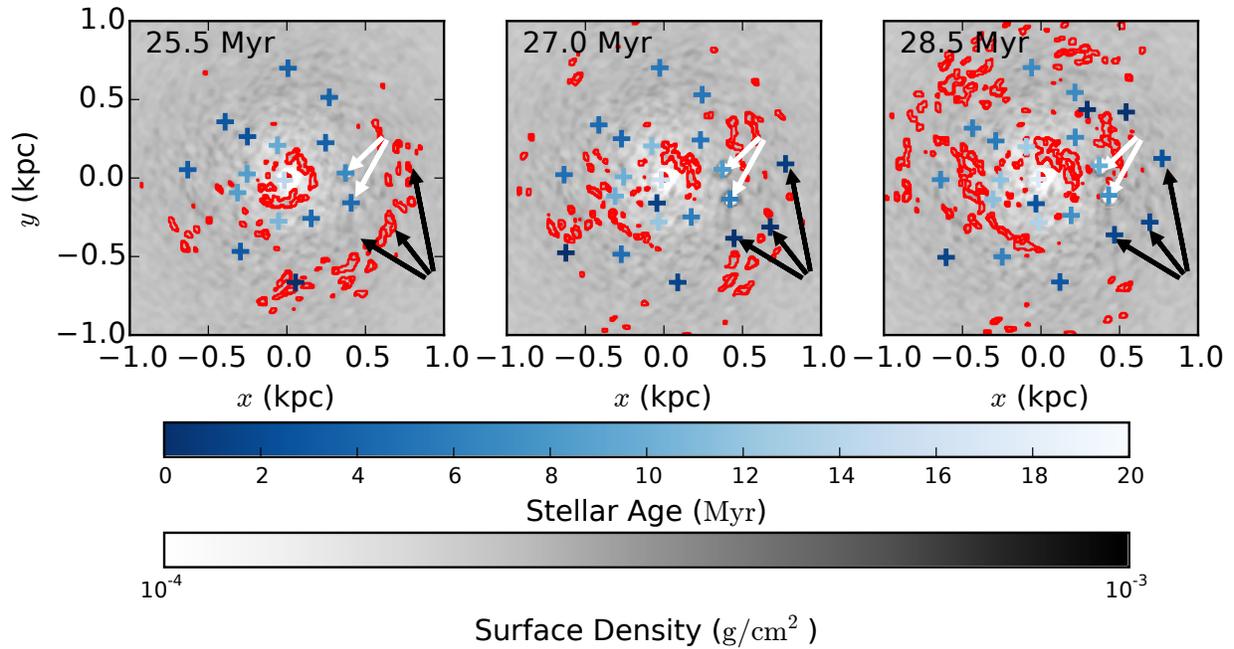

Figure 4. The effect of photoelectric heating. The surface density of gas is shown in grayscale, and the red contours show regions where the density is at least 80% of the Jeans density, i.e. the threshold density for forming stars at the gas's temperature. Star particles formed within the 20 Myr preceding each snapshot are shown as crosses, with darker colors indicating younger stars. The stars indicated with arrows, formed between the first and second snapshots, by the third snapshot have heated nearby gas preventing star formation. Supernovae remnants (white arrows) have clear morphological signatures, but do not affect star formation substantially.

**Methods**

Our simulations follow the evolution of an isolated dwarf galaxy for ~100 Myr using the AMR code Enzo[16]. We use a piecewise-parabolic mesh hydrodynamics solver, with an HLLC Riemann solver to follow the motion of the gas. The gravitational potential is computed on the same mesh used to solve the hydrodynamics. Dark matter and stars are included as collisionless particles acting independently of the hydrodynamics, except when new particles are formed or existing particles inject mass, metals, and energy back into the gas according to our feedback model.

We use the same sort of initialization as in previous work[28], though of course with parameters appropriate for a dwarf. In particular, we create a set of stellar and dark matter particles with the MakeGalaxy code[29]. The density of a stellar disk and bulge are pre-specified, and their velocities are set so that the galaxy begins in approximate Jeans equilibrium. The dark matter has an initial Hernquist profile. We use a halo mass of $10^{10}$ $M_\odot$, a concentration of 10, a spin parameter of 0.04 a stellar mass of $10^7$ M , and a stellar scale length of 300 pc.

We initialize the gas disk according to

$$\rho_d(r,z) = \rho_0 \exp\left(-\frac{r}{r_d}\right)\exp\left(-\frac{|z|}{H}\right)$$

This equation applies until the pressure $\rho_d T_d$ falls below the halo pressure $\rho_h T_h$, at which point the density and temperature are set to $\rho_h$ and $T_h$. While $\rho_d$ is spatially-dependent, $\rho_h$, $T_d$, and $T_h$ are all taken to be constant. In these simulations we set $T_h=10^6$ K, $T_d=1000$ K, $\rho_h=2.34 \times 10^{-30}$ g/cm$^3$, and $\rho_0 = 1.41 \times 10^{-24}$ g/cm$^3$. The scale length and scale height are $r_d = 5$ kpc and $H = 30$ pc. These initial conditions are chosen to minimize the time the disk spends in its initial collapse phase, and minimize the influence of the galactic halo on the dynamics. In particular, at this temperature the halo does not monolithically cool onto the galaxy over the timescale on which the simulation is run.

These initial conditions are evolved under the influence of hydrodynamics, gravity, and cooling, with additional subgrid prescriptions for the creation of new star particles and feedback from young stars, specifically stellar winds, type II SNe, and a rough treatment of heating from photoionization. We also include a new prescription for self-consistently calculating the volumetric heating rate from the grain photoelectric effect.

**Feedback** Our feedback prescription is designed to avoid the ad-hoc modifications to the physics that are typically necessary to produce feedback strong enough to affect the galaxy's properties. Our goal is to show that none of these modifications are necessary to have effective feedback at sufficiently high resolution. The basic physical picture is that once the adiabatic radius of the SNe can be resolved, they will expand in a somewhat realistic way, rather than immediately losing all their energy to the cell, as in the famous overcooling problem[22].

When each star particle is formed in the simulation with initial mass $M_p$, we draw a number from a Poisson distribution to determine how many type II SNe will be produced by the particle. The rate parameter of this Poisson distribution is taken to be $\lambda = 1104\, M_p/(10^6\, M_\odot)$, where the prefactor is the number of SNe produced over the lifetime of a $10^6\, M_\odot$ stellar population formed in a burst of star formation according to Starburst99[30] (SB99) using a Chabrier IMF[31].

For each SN produced, we next draw a number from the delay time distribution of type II SNe, again determined from the output of SB99. This is essentially just the convolution of the lifetime of massive stars with the IMF. At the hydrodynamical time step during which the SN explodes, we add $10^{51}$ ergs to the internal energy of the cell where the SN resides. The mass of ejecta and its metallicity are also determined from the SB99 output, and are fit to piecewise-polynomial functions for use in Enzo. To ensure that the supernova goes off on the highest level of refinement in the simulation, we make the star particles "must-refine" if the particle has any SNe remaining in its future. This means that the cells containing such particles are marked for refinement to the highest resolution level, guaranteeing that they and at least the surrounding two cells in each direction will be on the highest resolution level.

In addition to the energy from the SNe added to the cell at the appropriate time step, we also add energy to the cell prior to the supernova itself. From the delay time for each SNe, we infer the mass of the star that will be exploding, from which we can estimate the total ionizing luminosity for the particle by adding up the contribution from every massive star that has yet to explode in that particle. Given an ambient density, i.e. the density in the surrounding gas cell, we may then estimate the volume of gas which can be ionized by the calculated luminosity, as in the standard Strömgren analysis. If this volume is larger than the cell, and the cell has a temperature below $10^4$ K, internal energy is added to the cell to set the temperature to $10^4$ K. The cell is not modified by this photoionization prescription if its temperature is above $10^4$ K.

For cells with Strömgren volumes smaller than the cell volume, we compute the internal energy of the Strömgren sphere and compare it to the internal energy across the entire volume of the cell. If the whole cell has less internal energy than the Strömgren sphere alone, we add internal energy to the cell to make the two equal, and otherwise we do nothing.

This treatment of photoionization feedback is extremely conservative. In the limit of small cell sizes or low densities, we may vastly underestimate the size of the HII region, since only one cell will be at $10^4$ K. In high-density or large cells, the total energy being injected into the gas is also conservative, in that the cell's energy is only changed if the entire internal energy of the cell is smaller than the internal energy associated with the Strömgren sphere on its own.

In addition to photoionization, we include feedback from stellar winds. Once again we rely on Starburst99 to compute the wind luminosity, with several small modifications. The specific energy of the wind is taken from Starburst99 only when the particle will in the future experience a SN from the death of a star with a delay time less than 10 Myr. If no such SNe will explode in the future, the cell still loses mass to winds, but their specific energy is set to $(10 \text{ km/s})^2$. This bifurcation in wind temperatures takes place in the real universe as a result of a sharp change in opacity in the atmospheres of stars as a function of mass.

One additional change we make to the specific energy of the wind occurs right at the beginning of the star particle's life. During this time the most massive stars in the population drive extremely hot winds, exceeding $10^9$ K. When the outflowing winds dominate the material in a single cell as sometimes happens in our simulations, this high temperature can slow down the simulation substantially and cause other numerical problems associated with a high density contrast. To ameliorate this issue we cap the wind specific energy at $10^{8.5}$ K. We keep the total wind energy injected constant by slightly increasing the mass lost during this time. This changes the total mass returned by a tiny amount since this phase is so short.

Extended Data Figure 1 shows that every supernova in the 10 pc resolution simulation with the 5 kpc initial scale length explodes in a cell with a density below the critical density at which a supernova remnant would cool before expanding to be the size of a single cell in the simulation. We calculate this critical density by adopting the following value for the radius at which a supernova remnant exits the Sedov phase and enters the pressure-driven snowplow (PDS) phase[32],

$$R_{PDS} = 14 \frac{E_{51}^{2/7}}{n_H^{3/7}(Z/Z_{sol})^{1/7}} \text{ pc}$$

We take the energy of the supernova in units of $10^{51}$ ergs to be $E_{51}=1$, and set the mass fraction of elements heavier than helium to a value appropriate for the sun, $Z=Z_\odot=0.02$. In our initial conditions, the disk component has $Z/Z_\odot=0.1$, and the halo has $Z/Z_\odot=0.01$, but for the purposes of calculating the critical density we use a higher value because supernovae produce enough metals that locally $Z$ may be substantially higher than its initial value. By setting $R_{PDS}$ to the size of a single cell in the simulation, we can solve for the value of $n_H$, the number density of hydrogen atoms in units of cm$^{-3}$, at which the PDS phase would be marginally resolved.

By resolving this crucial piece of physics in these galaxies, we find that our results are relatively insensitive to the resolution at which we run the simulations. We compare the depletion time for each physical scenario, run at 10 pc, 5 pc, and 2.5 pc resolution, in Extended Data Fig. 3, and we find that the results of the simulations tend to become independent of resolution after roughly 100 Myr of evolution. This is less clear for the runs which include SNe, for which the 2.5 pc simulations have not advanced as far as their lower-resolution counterparts, but even here there is reasonable agreement between the 10 pc and 5 pc runs.

**Photoelectric Heating** FUV photons from young stars liberate electrons from dust grains in the interstellar medium. This is the primary means by which the neutral atomic gas in the interstellar medium is heated in the Milky Way. To include it in our simulation, we assume the following proportionality.

$$\Phi \propto F_{FUV} Z n_H \text{ [erg/s/cm}^3\text{]}$$

The heating rate from FUV photons is proportional to their flux, and the density of metals. At low densities and high temperatures, there is an additional dependence on the electron density and gas temperature, but these effects are negligible in the cold, dense gas where FUV heating is important for suppression of star formation, so we omit that effect. We also do not include cosmic ray heating, since this is roughly an order of magnitude less important than FUV heating under the optically thin conditions that prevail in the low density, dust poor galaxies we are simulating.

We calculate the FUV flux in the simulation by taking the luminosity for each star particle to be

$$\log_{10} L(t_7) = \log_{10}(M_p / 10^6 M_{sol}) + \begin{cases} \sum_{j=0}^{2} p_j t_7^j & \text{if } t_7 \leq 3 \\ \sum_{j=0}^{6} q_j t_7^j & \text{if } 3 \leq t_7 \leq 10 \\ 0 & \text{otherwise} \end{cases}$$

where $t_7$ is the age of the star particle in units of 10 million years, and $M_p$ is the mass of the star particle. $L$ is in units of erg/s. The coefficients of this polynomial equation are given in Extended Data Table 1. This is the result of integrating the output spectrum of a Starburst99 single-burst model for a cluster mass of $10^6$ M$_{sol}$ (hence the pre factor in the previous equation) over the range 8 to 13.6 eV at finely spaced time intervals out to 100 Myr. Note that this function is somewhat sensitive to the IMF - recent indications of a bottom-light IMF[33] in dwarfs would somewhat increase it by a factor of ~2.

The FUV flux is then simply $F_{FUV} = L(t)/(4\pi r^2)$, where r is the three-dimensional distance from the center of the gas cell in question to the star in question. Note that this neglects any effects from self-shielding, which should be negligible in the galaxies we have simulated, given their low column densities and low metallicities. The total FUV flux at a given cell is the sum of this quantity over all stars in the simulation. If a given star's contribution to the FUV flux varies by less than 10% across a given grid (the computational element one step above cells in Enzo), then we approximate that star's contribution as constant across the grid, to avoid doing the full order N by M computation (where N is the number of cells and M is the number of particles) in regions far from FUV-emitting particles. To compute $\Phi/n_H$, we scale $\Phi/n_H$, $F_{FUV}$ and $Z$ to the known values of these quantities in the solar neighborhood,

$$\frac{\Phi}{n_H} = 8.5 \times 10^{-26} \frac{\text{erg}}{\text{s}} \frac{F_{FUV}}{0.0015859021 \text{ erg/s/cm}^2} \frac{Z}{Z_{sol}}$$

The numerical constant in the denominator is simply the Habing[34] estimate of the intensity of the interstellar radiation field in the solar neighborhood, multiplied by $c$ to convert to a flux. The pre factor of $8.5 \times 10^{-26}$ is the photoelectric heating rate normalized to the Habing value for the solar neighborhood from a radially-dependent model of the Milky Way's ISM[35]. This quantity is computed for each cell in the simulation, and fed to the Grackle library, which computes the rate of change of the internal energy density.[16,36,37] An example of $\Phi/n_H$ in the PE Only simulation is shown in Figure 3.

This method is a substantial improvement over many current implementations of the interstellar radiation field (ISRF). Many simulations do not include this source of diffuse heating at all. Some include it as constant throughout the simulation volume[3,4], sometimes with a correction for self-shielding[5], but without regard to the individual sources or time-dependence of the FUV photons. Other simulations explicitly approximate the radiative transfer of these photons[38], while others go even further and explicitly model the dust particles via which this radiation interacts with the gas[2,39,40]. Although the latter two methods are better approximations to the ISRF than what we have implemented here, our simulations have a higher resolution. As a result, we resolve the Sedov-Taylor phase of the SNe (Extended Data Fig. 1), and hence can definitively show whether supernova feedback or the ISRF is dominant in the regulation of star formation in dwarf galaxies.

**Star Formation** The star formation prescription we use is similar to many commonly-used schemes. At each time step the probability of forming a star in each cell is taken to be

$$p_* = \begin{cases} 0 & \text{if excluded} \\ \min\left(1, \varepsilon_{ff} \frac{dt}{t_{ff}} \frac{M_{cell}}{M_p}\right) & \text{otherwise} \end{cases}$$

where the simulation time step (at the refinement level in question) is $dt$, $\varepsilon_{ff}$ is the efficiency of star formation per free fall time, and the free fall time $t_{ff} = \sqrt{(3\pi/32G\rho)}$. Cells are excluded if their mass is less than the Jeans mass in that cell or if the cell is not on the maximum refinement level. In other words, cells must have a density exceeding

$$\rho > \frac{\gamma \pi k_B T}{N_J^2 G \mu m_H (\Delta x)^2}$$

in order to form stars. Here $\gamma=5/3$ is the ratio of specific heats, $\mu$ is the mean molecular weight in units of the Hydrogen mass, $\Delta x$ is the size of the cells on the maximum refinement level, and $N_J=4$ is the number of cells by which we require that the Jeans length be resolved throughout the simulation to avoid artificial fragmentation[41]. This criterion corresponds to a diagonal line in the density-temperature phase diagram above which stars are not allowed to form (Extended Data Fig. 4). The value of $p_*$ is such that the average star formation rate in the non-excluded cells follows a volumetric Schmidt Law.

$$\dot{\rho} = \varepsilon_{ff} \rho / t_{ff}$$

This model has a few parameters that we must set, although we argue that we do not have a huge amount of freedom to change them.

The efficiency per free fall time is constrained by observations to be within a factor of a few of $0.01$[42,43], so we simply adopt this central value. The mass of individual star particles $M_p$ should in principle be low enough to be irrelevant -- this would have the advantage of sampling the star formation rate density very well and forming stars if and only if the cell were Jeans unstable. This introduces two numerical difficulties. The first is that the Jeans mass for the coldest gas in our simulations can be comparable to the mass of a single star. It would be problematic for our feedback recipe to form such low-mass stars, in that a 120 solar mass star or a supernova which ejects 10 solar masses of material could never exist within a 10 solar mass particle. Moreover the number of star particles we would have to follow increases as $M_p^{-1}$. Each of these issues would be avoided by choosing a large particle mass. However, using a large mass increases the chance that the cell in which the star formed would have insufficient mass to supply all of the gas needed to form the particle. When this happens, in order to maintain global mass conservation we gather the mass from neighboring cells. Each cell supplying gas contributes the same fraction of its mass to the new particle. We have found that a particle mass of $M_p = 50\ M_\odot$ strikes a reasonable balance between keeping star formation local to 1 cell and keeping the supernova ejecta mass less than a single star particle's mass.

**Comparison to observations** In Extended Data Figure 2, we compare the star formation rate and depletion time of the simulations to those of dwarf galaxies from three different samples[14,20,24]. We exclude four blue compact dwarf galaxies from this sample, since these are known to be undergoing starbursts. We do not include estimates for the errors on each point. The statistical standard errors are of order 10%, but this under-predicts the true error, which is dominated by systematic uncertainty in distance (HI mass and star formation rate) and the assumptions made in converting UV or Hα luminosities to star formation rates[44]. For the two smaller galaxies in ref. 20, no strict upper limits are given on the star formation rate, but since they are not detected in the same Hα image in which an SFR of $4 \times 10^{-3}$ $M_{sol}$/yr was measured for the largest galaxy in the sample, we take this as a very conservative upper limit on the SFR for these two galaxies (Janowiecki, priv. comm.).

**Data Availability** The numerical experiments presented in this work were run with a fork of the enzo code available from https://bitbucket.org/jforbes/enzo-dev-jforbes, in particular change set daed04d1e5e6. This altered version of enzo also requires an altered version of the grackle cooling library, available from https://bitbucket.org/jforbes/grackle, particularly change set 12d3856. A subset of the raw data files are also available online from www.johncforbes.com/dwarfs.html.

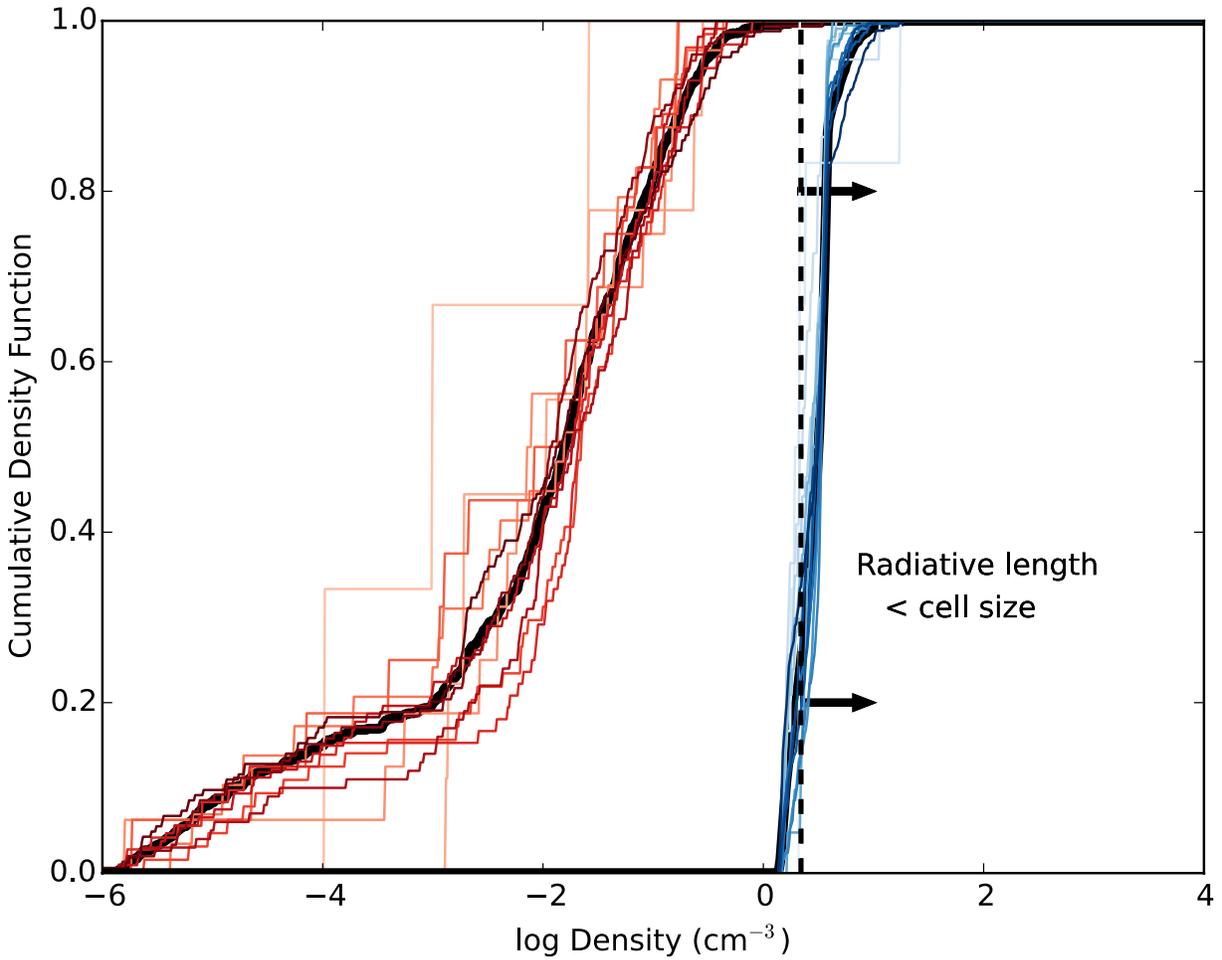

Extended Data Figure 1. Supernovae are well-resolved. The cumulative distribution of the density of cells in which supernovae explode is shown in red, and in which stars form is shown in blue, with the thin lines showing these distributions in different 10 Myr intervals. The vertical dashed line indicates the density at which a supernova remnant would radiate all of its energy before it expanded to the size of a single cell (10 pc) in the simulation, assuming solar metallicity. Nearly every supernova in the simulation occurs to the left of this line, indicating that the simulation does not suffer from the overcooling problem.

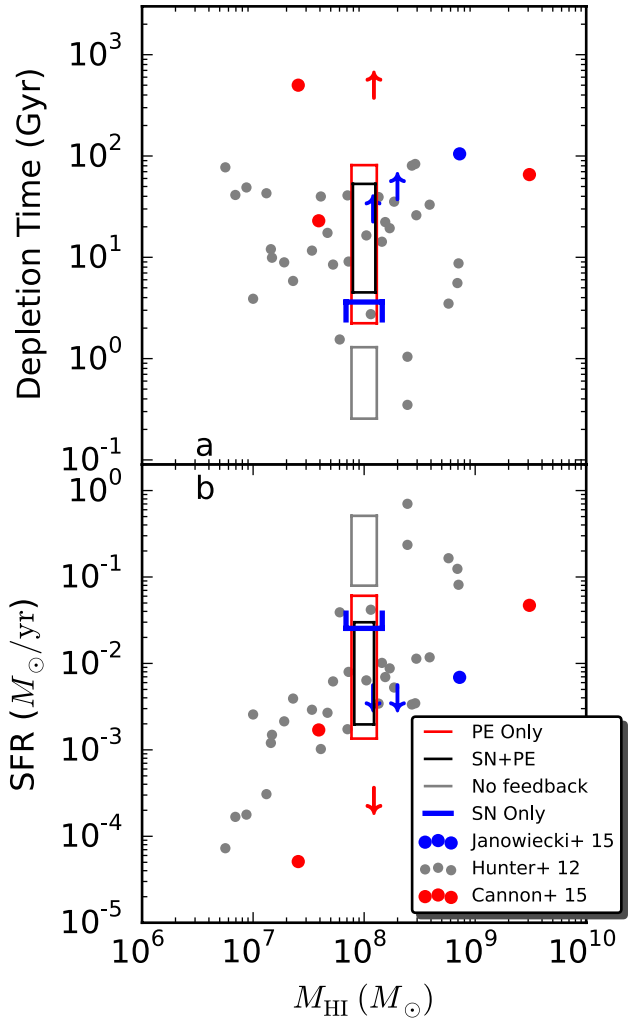

Extended Data Figure 2. Comparison to observations. Star formation properties for a heterogeneous set[14,20,24] of galaxies are shown as a function of gas mass. Boxes representing the range of values covered by our simulations are overplotted. The vertical range of the box is determined by the final snapshot for each simulation; the high-(low-) SFR extremum represents the 1 (5) kpc disks. No "SN Only" simulation was run for the 1 kpc case, so the blue box is not closed. Only simulations that include photoelectric heating agree with the depletion times observed for bulk of galaxies in the mass range we simulated.

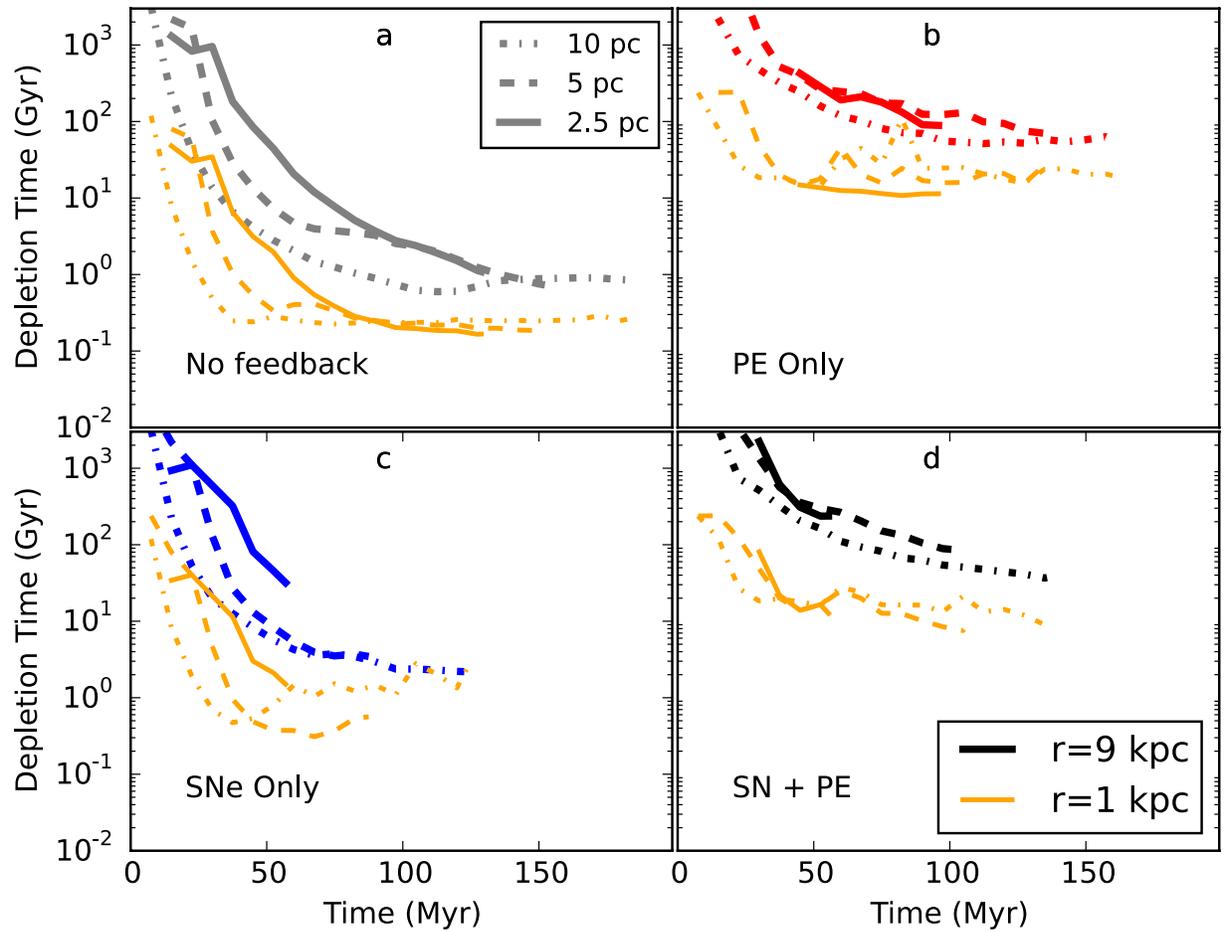

Extended Data Figure 3. A resolution study. The depletion time of all 12 simulations with 5 kpc gas scale length is plotted over time. The panels show the four different feedback models, while the lines of a given color in each plot show the result for different resolutions. The orange lines show the depletion time when the measurement is carried out in a cylinder with 1 kpc radius, whereas the other lines use a 9 kpc radius. Regardless of the aperture, the simulations quickly converge; differences between simulations with factor-of-two differences in resolution are small compared to the differences resulting from changing the physics.

| j | $p_j$ | $q_j$ |
|---|---|---|
| 0 | 41.5709926107 | 40.7875024388 |
| 1 | 2.40501751872 | -0.227682606645 |
| 2 | -9.19544984847 | 0.0078916423535 |
| 3 | 10.5203892767 | 0 |
| 4 | -5.72637964222 | 0 |
| 5 | 1.50479521662 | 0 |
| 6 | -0.153355377095 | 0 |

Extended Data Table 1. Parameters for a fit. The piecewise polynomial fit to the FUV luminosity of a simple stellar population as a function of its age (Methods section, equation 4) uses these coefficients.

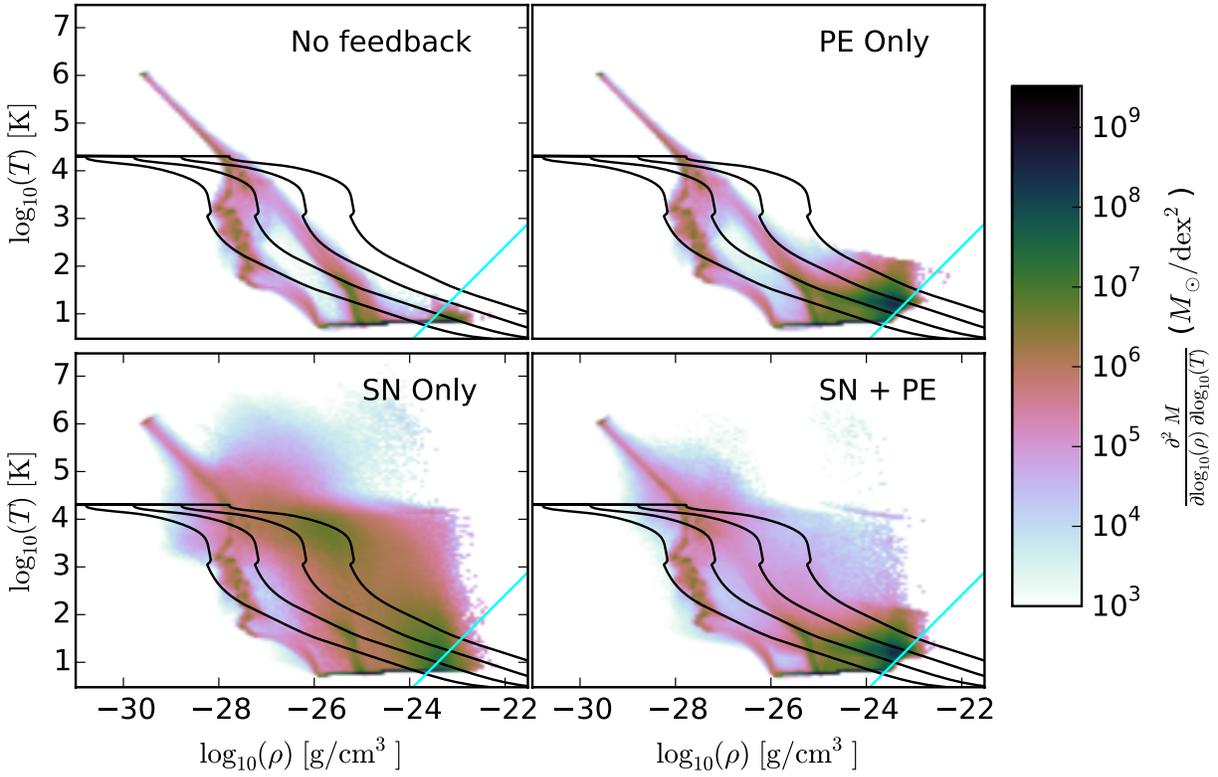

Extended Data Figure 4. Phase diagrams after 90 Myr of evolution. The panels show runs with different feedback models and 5 kpc gas scale length, all at 10 pc resolution. The light blue diagonal lines show the threshold for star formation, namely where the gas becomes Jeans unstable on the highest refinement level. The black lines trace where the net cooling rate is zero, assuming different values for the volumetric heating rate, from $10^{-26}$ erg/s (highest line), to $10^{-29}$ erg/s. Photoelectric heating raises the typical temperature of gas near the star formation threshold such that moderate star formation can stabilize nearby gas against collapse.